\documentclass[12pt]{article}

\usepackage[a4paper, total={6in, 8in}]{geometry}
\usepackage[pdftex]{graphicx}
\usepackage{caption}
\usepackage{natbib}

\usepackage{mathtools}
\usepackage{amsmath}
\usepackage{amssymb}

\usepackage[dvipsnames]{xcolor}
\usepackage{cleveref}

\usepackage{authblk}

\usepackage{tikz}
\usetikzlibrary{decorations.pathreplacing,angles,quotes}

\DeclareRobustCommand\sampleline[1]{%
  \tikz\draw[#1] (0,0) (0,\the\dimexpr\fontdimen22\textfont2\relax)
  -- (2em,\the\dimexpr\fontdimen22\textfont2\relax);%
}

\DeclareFontFamily{U}{solomos}{}
\DeclareErrorFont{U}{solomos}{m}{n}{10}
\DeclareFontShape{U}{solomos}{m}{n}{
  <-> s*[1.1]  gsolomos8r
}{}

\title{Transitional Atmospheric Boundary Layer in the GABLS4 Experiment Modelled Using the Explicit Algebraic Reynolds-stress Model}

\author[1]{Velibor \v{Z}eli}
\author[1]{Stefan Wallin}
\author[1]{Arne V.~Johansson}
\author[1]{Geert Brethouwer}

\affil[1]{Department of Engineering Mechanics, FLOW Centre, KTH Royal Institute of Technology, 10044 Stockholm, Sweden}

\date{}

\graphicspath{{./}{./figures/}}

\begin{document}

\maketitle

\begin{abstract}
    A recently developed so-called explicit algebraic Reynolds-stress (EARS) model is applied to a transitioning atmospheric boundary layer (ABL). The simulation describes a diurnal cycle with a deep convective ABL during daytime and an extremely thin and stably stratified ABL during nighttime. The predictions of the EARS model are compared to large-eddy simulations (LES) of Couvreux \emph{et al.}~ (Bound Layer Meteorol 176:369-400, 2020). The model simulation is extended in time in order to study several consecutive diurnal cycles. The EARS model uses the same parametrization and model coefficients for stable and convective ABL and is applicable over a wide range of thermal stratifications. First-order statistics are shown to be well predicted by the model. We also show that the model can predict transitional effects such as residual turbulence as well as horizontal turbulent fluxes, which are an inherent part of the EARS model solution.
\end{abstract}

\section{Introduction}

Atmospheric boundary layers (ABLs) developing over land are in a state of continuous change and determine the exchange of momentum and heat between the surface and overlying air. The ABL is generally turbulent with a pronounced diurnal cycle. During daytime, when the surface gets warmer, the ABL is often deep and convective while during nighttime the surface cools and the ABL is shallow and stably stratified. In between these states of distinct stratification, the ABL undergoes morning and afternoon/evening transitions. These directly impact the diurnal cycle of the near-surface variables, e.g., the low-level jet \citep{smith2019great} is a result of evening transition in ABLs and  has an impact on severe weather and also strongly affects synoptic-scale systems. Therefore, modelling of transitioning effects in the ABL is an essential part of atmospheric modelling \citep{angevine2020transition, edwards2020representation}.

To evaluate turbulence models intercomparison studies have been organized in the Global Energy and Water Exchanges (GEWEX) Atmospheric Boundary Layer Study (GABLS) \citep{holtslag2013stable}. Large-eddy simulations (LES) have been performed for an idealized geometry as well as atmospheric observations. The capability of turbulence models is tested by comparisons with these observations and LES. So far there have been four GABLS experiments, each designed to asses turbulence models for specific atmospheric conditions. The latest so-called GABLS4 experiment by \cite{couvreux2020intercomparison} has multiple LES contributions and provides a benchmark for modelling of a diurnal cycle with relatively deep convective ABL during daytime and an extremely thin ABL with weak geostrophic forcing and high stability during nighttime.

The explicit algebraic Reynolds-stress (EARS) model by \cite{lazeroms2013explicit} (based on the modelling in \cite{wallin2000explicit} and \cite{wikstrom2000derivation}) was validated by \cite{vzeli2020modelling} for the case of a modified GABLS1 experiment by comparing model results to LES by \cite{sullivan2016turbulent}. The results show that the EARS model is able to predict first- and second-order statistics in ABL with good accuracy for different levels of stable stratification. \cite{vzeli2020explicit} applied the same formulation of the EARS model to simulate a convective ABL and compared the results to LES by \cite{salesky2018buoyancy}. These studies have shown that the EARS model can be successfully used for predicting turbulent ABLs for different degrees of thermal stability ranging from stable to convective. This makes the EARS model different from turbulence models with disparate descriptions for stable and unstable stratifications, and even for different levels of stratification (see \cite{he2019new}). \cite{vzeli2019consistent} used the EARS model for simulating the diurnal cycle from GABLS2 by \cite{svensson2011evaluation}. The results were compared only to the Monin--Obukhov theory inside the surface layer. This study demonstrated the validity of boundary-conditions treatment implemented in the EARS model. \cite{lazeroms2016study} studied a diurnal cycle in a dry ABL with harmonic temperature at the surface and strong geostropic wind forcing. They report that the EARS model is able to capture transitional effects such as residual turbulence, but they had no data for comparison.

The purpose of this study is to evaluate the EARS model in modelling transitioning processes in an ABL. The EARS model is used for simulating the diurnal cycle in GABLS4 and compare the results to several LES by \cite{couvreux2020intercomparison}. The study is based on measurements by \cite{genthon2013two} taken at Dome C in Antarctica. The LES case represents an ideal and dry diurnal cycle with weak geostropic forcing. A strong surface cooling rate makes prediction of the transition from deep convective to strongly stratified ABL challenging.

The following section gives a case description of GABLS4. Section \ref{sec:results} describes the results of the EARS model prediction. Conclusions are given in Sect. \ref{sec:conclusion}.

\section{Case Description} \label{sec:case}

Results of the EARS model are compared with the so-called GABLS4 LES study by \cite{couvreux2020intercomparison}. Namely, we focus on three particular LES from that study: DALES, MicroHH and PALM due to their mutual consistency. These LES form a band where the width of the band is the spread between the results.

\begin{figure}
  \centering
  \includegraphics{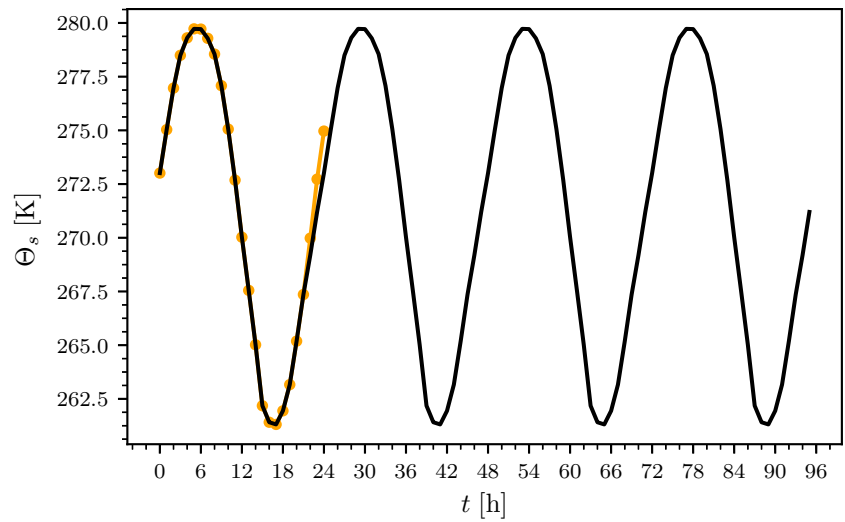}
  \caption{Potential temperature at the surface as a function of time for the original LES (orange) and the extended simulation with modified boundary conditions in order to achieve smooth periodicity (black). The temperature forcing at the surface drives the ABL through four identical diurnal cycles in which the ABL transitions between unstable and stable thermal stratification}
  \label{fig:Theta_s}
\end{figure}

A dry ABL is driven by prescribed time-varying potential temperature at the surface $\Theta_s$ and constant geostrophic wind $\mathbb{V}_g = (U_g = 1.25, V_g = 4.5) \, \mathrm{m} \, \mathrm{s}^{-1}$ along the x- and y-directions. The simulations are preformed for an elevation of $3233 \; \mathrm{m}$ where the atmospheric pressure is constant in time $p_a = 651 \; \textrm{hPa}$. The potential temperature at the elevated surface corresponds to the surface temperature forcing from \cite{couvreux2020intercomparison} and is expressed as

\begin{equation}
  \Theta_S = T \left(\frac{1000}{p_a}\right)^{0.286},
\end{equation}

\noindent where $T$ is the surface temperature from \cite{couvreux2020intercomparison}. The GABLS4 experiment studies a single diurnal cycle starting with morning transitioning into convective and developing into a stably stratified ABL in the afternoon. The focus of the experiment is to investigate a horizontally homogeneous transitioning ABL with strong surface cooling rate. The original surface boundary conditions are nearly but not fully periodic in time. To be able to continue the model simulations we have therefore extended the LES setup by slightly modifying the temperature forcing for the final three hours of the simulation such that the surface temperature is smooth and periodic in time. The periodic lower-boundary condition allows us to extend the original GABLS4 experiment to four consecutive days. Figure \ref{fig:Theta_s} shows the prescribed surface temperature as a function of time in the original GABLS4 experiment and the modified periodic extension used in the present study. The initial 22 hours in the GABLS4 and current study are identical, which ensures a valid comparison of the results from the EARS model and LES during that period. Because the simulation is extended to four days we can study the transition from stable to convective ABL. Having a longer simulation also reduces the influence of initial conditions on the overall result by allowing TKE and surface fluxes to reach a nearly periodic state.

The ABL is developing over a plane surface. The domain is bounded between the roughness height for momentum $z_{0_m} = 0.01 \, \mathrm{m}$ and $H = 1000 \, \mathrm{m}$ with 121 grid points log-linearly distributed up to $500 \; \mathrm{m}$ and from there equidistantly to the upper boundary. Other simulation parameters are roughness height for temperature $z_{0_h} = 0.001 \, \mathrm{m}$, Coriolis parameter $f = -1.4 \times 10^{-4} \, \mathrm{s}^{-1}$, reference temperature $T_0 = 265 \; \mathrm{K}$, and gravitational acceleration $g = 9.81 \, \mathrm{m} \, \mathrm{s}^{-2}$.

\begin{figure}
  \centering
  \includegraphics{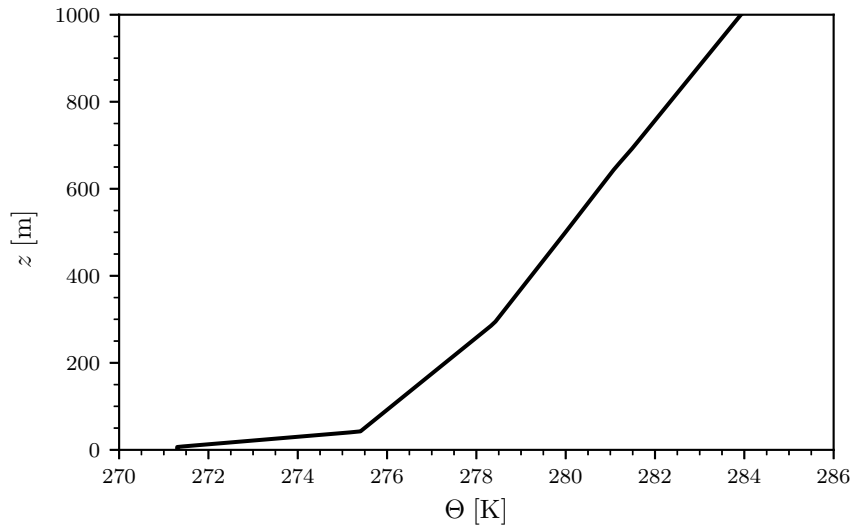}
  \caption{Mean potential temperature profile used for initialization}
  \label{fig:Theta_0}
\end{figure}

The initial wind profile in x- and y-direction is equal to the geostrophic wind. The initial state represents a stable stratification with a mean potential temperature profile shown in Fig.~\ref{fig:Theta_0}. The initial TKE profile is $K = 0.4(1-z/250)^3 \, \mathrm{m}^2 \, \mathrm{s}^{-2}$ for $z < 250$ $\mathrm{m}$. Above $z = 250 \, \mathrm{m}$ $K$ is equal to $10^{-5} \, \mathrm{m}^2 \, \mathrm{s}^{-2}$. The initial profile for the TKE dissipation rate $\epsilon$ is prescribed through specifying the turbulence time scale $\tau_0 = K/\epsilon = 1 \, \mathrm{s}$ at $z_0$ increasing linearly to $\tau_0 = 550 \, \mathrm{s}$ at 250 m, above which it is constant. This means that the turbulence eddies close to the surface are small, and the size of the turbulence structures increases with height. The initial profile for the half potential temperature variance $K_\theta = \frac{1}{2} \overline{\theta^2}$ is set to $10^{-4} \, \mathrm{K}^2$, resulting in small levels of potential temperature variance $\overline{\theta^2}$. The effect of the initial conditions for $K$, $\epsilon$, and $K_\theta$ on the model results is negligible compared to the effect of the initial condition for $\mathrm{\Theta}$.

The EARS model is implemented in the context of a single-column model with spatial variation only in the vertical direction. The complete model description and boundary-condition treatment are given in \cite{vzeli2019consistent}. Model coefficients are the same as the coefficients used for stable and convective ABL in \cite{vzeli2020modelling} and \cite{vzeli2020explicit}, respectively. The complete set of the coefficients is $C_{\epsilon 1} = 1.44,\, C_{\epsilon 2} = 1.82,$ $C_{\epsilon 3} = 0$, $\sigma_K = 1$, $\sigma_\epsilon = 1.3$, $\sigma_{K_\theta} = 1$, $c_1 = 1.8,\,c_2 = 5/9,\,c_3 = 0.35,\,c_{\theta 1} = 4.51,\,c_{\theta 1}^* = 0.5,\,c_{\theta 2} = 1.0,$ and $c_{\theta g} = 1.0$. The model is implemented in the symbolic programming language Maple. Machine-generated Fortran code is used for obtaining the numerical solution of the time dependent one-dimensional problem by second-order differences in space and Crank--Nicolson in time.

\section{Results} \label{sec:results}

Results of the EARS model are compared to the LES by \cite{couvreux2020intercomparison}. First order statistics are compared at $t = 5 \, \mathrm{h}$ when the ABL is convective and $t = 17 \, \mathrm{h}$ when the ABL is stable. Here, $t$ is the time after the initialization of the LES and model simulations. These profiles are taken when the surface temperature in the LES and model simulations are the same ensuring a valid comparison. The mean profiles from the LES are time averaged for one hour.

\begin{figure}
  \centering
  \includegraphics{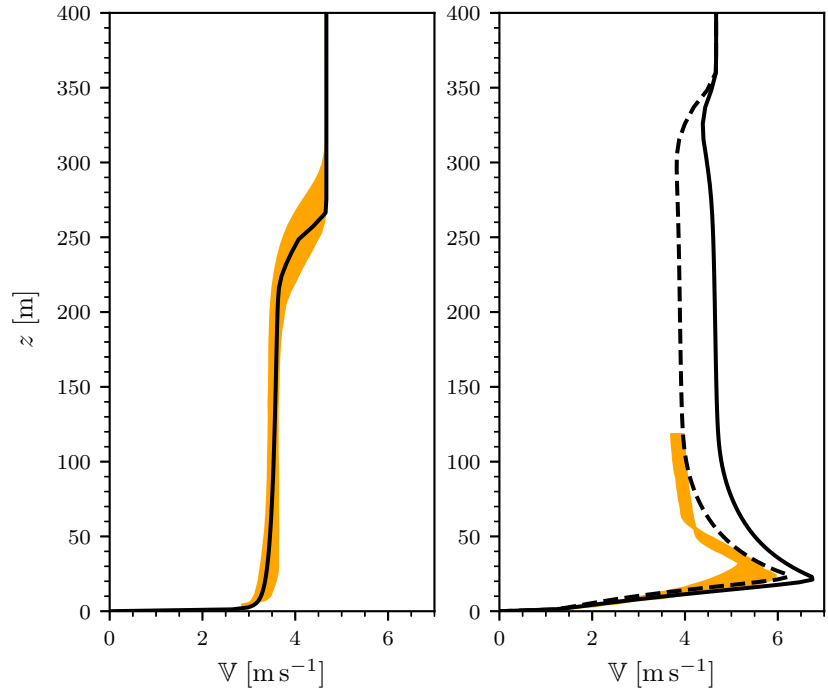}
  \caption{Solid lines are mean horizontal wind speed at $t = 5 \; \mathrm{h}$ (left) and $t = 17 \; \mathrm{h}$ (right) corresponding to convective and stable periods. Additionally, the dashed line is mean horizontal wind speed at $t = 19 \; \mathrm{h}$. Results of the EARS model are represented by the black line and the orange band is range of values of three LES}
  \label{fig:V}
\end{figure}

Figure \ref{fig:V} shows the mean profiles of horizontal wind speed $\mathbb{V} = \sqrt{U^2 + V^2}$, where $U$ and $V$ are the horizontal wind speed in the direction of x- and y-axis, respectively. The horizontal wind speed is fairly constant but lower than the geostrophic wind speed throughout the mixed layer. During the stable part of the diurnal cycle a low-level jet (LLJ) develops. The LLJ is commonly found in ABL during the period when the surface is being cooled (see \cite{beare2006intercomparison} and \cite{sullivan2016turbulent}). At $t = 17 \; \mathrm{h}$ the LLJ in the EARS model is nearer the surface and stronger than in the LES. This is opposite to the observation made in \cite{vzeli2020modelling} where the EARS model overestimates the height of LLJ. It could be related to the low grid resolution of the LES. \cite{sullivan2016turbulent} have shown that the LLJ lowers and strengthens with increase of grid resolution in LES. The maximum wind speed of the LLJ is 1.4 times the geostrophic wind, which is somewhat higher than 1.3 times the geostrophic wind reported in \cite{davis2000development}. The minimum wind speed above the LLJ is around 0.9 time the geostrophic wind while \cite{davis2000development} reported a value 0.6 times the geostrophic wind. Note that the mean wind profile at $t = 19 \; \mathrm{h}$ in the model simulations agree better with the LES. At that time the maximum of the LLJ is 1.3 and minimum is 0.8 times the geostrophic wind speed. This indicates that the difference might be partly caused by a difference in the initial transient in LES and model simulations.

\begin{figure}
  \centering
  \includegraphics{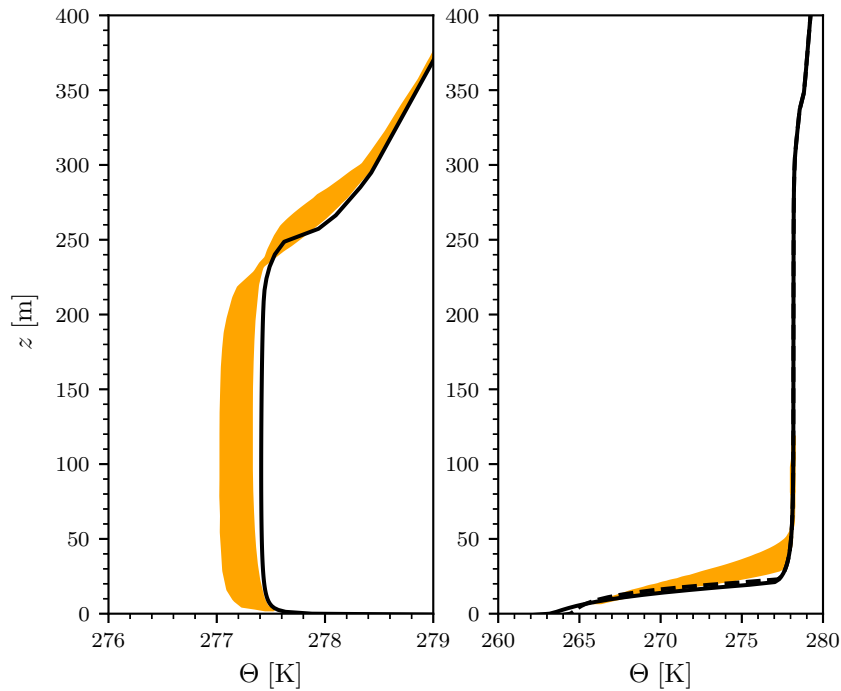}
  \caption{Mean potential temperature at $t = 5 \; \mathrm{h}$ (left) and $t = 17 \; \mathrm{h}$ (right) corresponding to convective and stable periods. Additionally, the dashed line is mean potential temperature at $t = 19 \; \mathrm{h}$. Results of the EARS model are represented by the black line and the orange band is range of values of three LES at $t = 17 \; \mathrm{h}$}
  \label{fig:T}
\end{figure}

Figure \ref{fig:T} shows the mean potential temperature profiles during the convective and stable ABL for LES and EARS model. For the convective ABL the mean potential temperature in the EARS model is higher than in the LES. However, the difference is less than one degree Kelvin. The gradient of the mean potential temperature is negative in the surface layer and lower part of the mixed layer while it is positive in the higher part of the mixed layer and entrainment zone. Similar observations were reported in \cite{vzeli2020explicit}, where the positive temperature gradient in the higher part of ABL was shown to be associated with the existence of a non-gradient term in the turbulent heat flux. In the stable ABL at $t = 17 \; \mathrm{h}$ the mean potential temperature monotonically increases. The height of the ABL is slightly lower than in the LES. This differs from the results reported in \cite{vzeli2020modelling}, where the EARS model predicted a higher ABL than the LES by \cite{sullivan2016turbulent}. \cite{sullivan2016turbulent} also reported that the stable ABL height becomes lower when refining the grid resolution. The LES in \cite{couvreux2020intercomparison} have a relatively coarse grid size of $2 \; \mathrm{m}$ while \cite{sullivan2016turbulent} used  0.39 m in the LES with the highest resolution for similar stable conditions. This could mean that the ABL height reduces when the grid is refined in the LES by \cite{couvreux2020intercomparison}.

\begin{figure}
  \centering
  \includegraphics{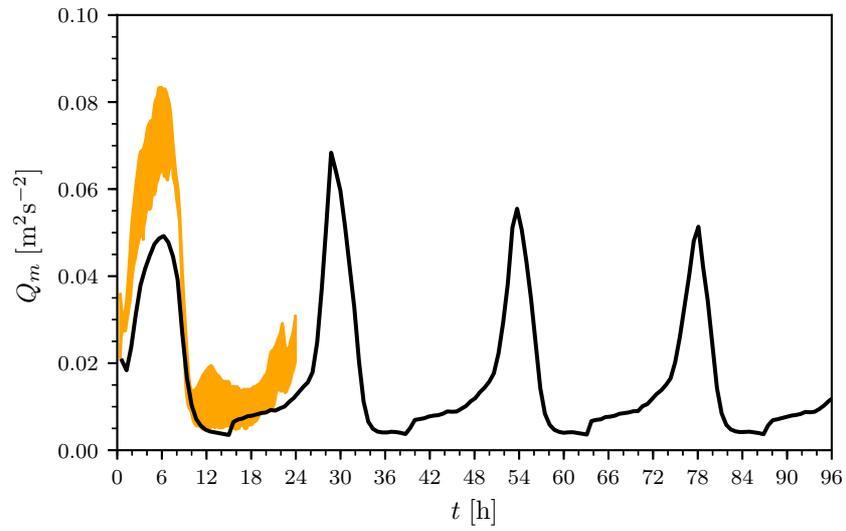}
  \caption{Surface  momentum flux as a function of time for the EARS (black line) and three LES models (yellow band)}
  \label{fig:Qm}
\end{figure}

\begin{figure}
  \centering
  \includegraphics{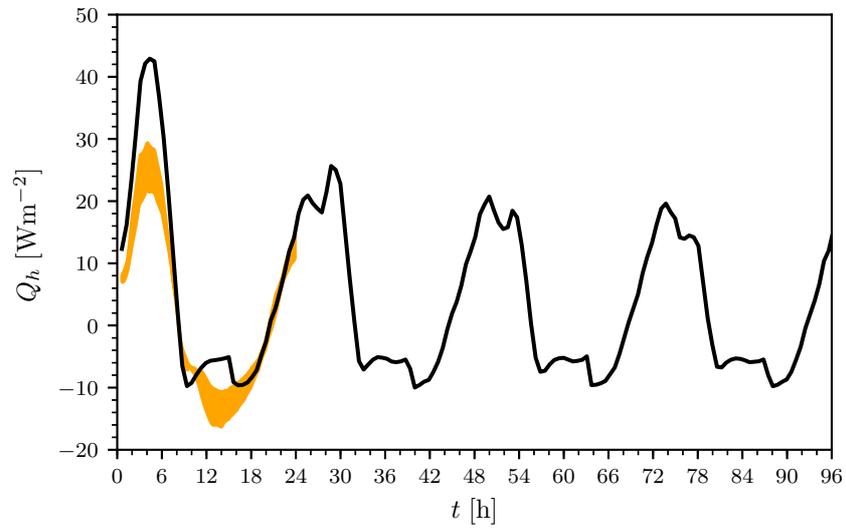}
  \caption{Surface heat flux as a function of time for the EARS (black line) and three LES models (yellow band)}
  \label{fig:Qh}
\end{figure}

Figures \ref{fig:Qm} and \ref{fig:Qh} show results of the surface momentum and heat fluxes as they evolve in the simulations. The surface momentum flux $Q_m = {u_*}^2$, where $u_*$ is the friction velocity, and the surface heat flux $Q_h =  \rho c_p u_* \theta_*$, where $\theta_*$ is the characteristic temperature for scaling in surface layer, $\rho = 1.225 \; \mathrm{kg} \, \mathrm{m}^{-3}$ is the air density and $c_p = 1003.5 \; \mathrm{J} \, \mathrm{K}^{-1} \, \mathrm{kg}^{-1}$ is the specific heat capacity at constant pressure. Both the surface momentum and heat flux in the EARS model deviate from the LES results in the initial eight hours of the simulation corresponding to the convective part of the day. These deviations are most likely related to the initialization of the simulations. It is interesting to note that the magnitude of the fluxes in the EARS model is significantly larger during the convective part of the first day than in the convective periods of the remaining days. This could indicate that the initial conditions play an important part in the setup of diurnal cycles. The magnitude of peaks in the third and fourth day indicate that the ABL is essentially independent of initial conditions in the third day. It would be interesting to extend the LES about twelve hours in time in order to capture one additional convective period less influenced by the initial conditions.

\begin{figure}
  \centering
  \includegraphics{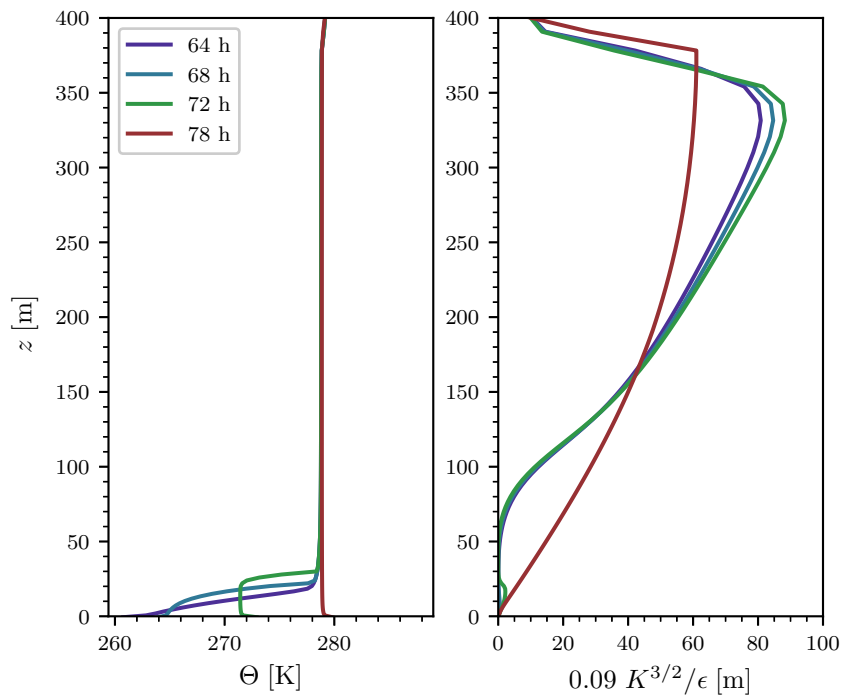}
  \caption{Profiles of mean potential temperature (left) and $0.09 K^{3/2} / \epsilon$, which can be interpreted as a qualitative estimate of the turbulence (macro-) length scale}
  \label{fig:l}
\end{figure}

Figure \ref{fig:Qh} shows that apart from the first day, which is strongly influenced by the initial conditions, there are two distinct moments during the convective ABL when the surface heat flux peaks. These peaks are a result of the transition from stable to convective conditions. To understand the process, we turn our focus to the fourth day where the first maximum appears around $t = 72 \mathrm{h}$ some time after the ground starts to heat the ABL. By following the detailed development of the temperature from $t = 64 \mathrm{h}$ in Fig.~\ref{fig:l}, we observe that a thin (30m) near-surface region develops with a nearly constant temperature around $t = 72 \mathrm{h}$ before the convection mixes the complete ABL at $t = 78 \mathrm{h}$. Figure \ref{fig:l} also shows a quantity that represents a qualitative measure of the turbulence length scale in a neutral ABL away from the immediate vicinity of the surface, where $\epsilon$ is the rate of dissipation of TKE. It shows that in the morning the air near the surface gets heated and that the turbulent length scale is quite small in the region up to about $60 \, \mathrm{m}$. This means that the convective ABL is not yet developed. Therefore, the first peak in the surface heat flux is related to the surface heating. The second peak appears at $t = 78 \mathrm{h}$ corresponding to the surface temperature maximum. At that time the convective ABL is fully developed and shows that the turbulent length scale is large, of the order of tens of meters, in the mixed layer, see Fig.~\ref{fig:l}. The turbulent mixing leads to the downward transport of warmer air, which was previously above the stable ABL. This mixing further increases the potential temperature in the surface layer and creates the second peak in the surface heat flux. This process results from ABL transitioning from a stably to unstably stratified state. Note that this process is observed in the second, third and fourth day but not in the first due to the influence of initial conditions. It is also not observed in the LES, which covers only the first convective period.

\begin{figure}
  \centering
  \includegraphics{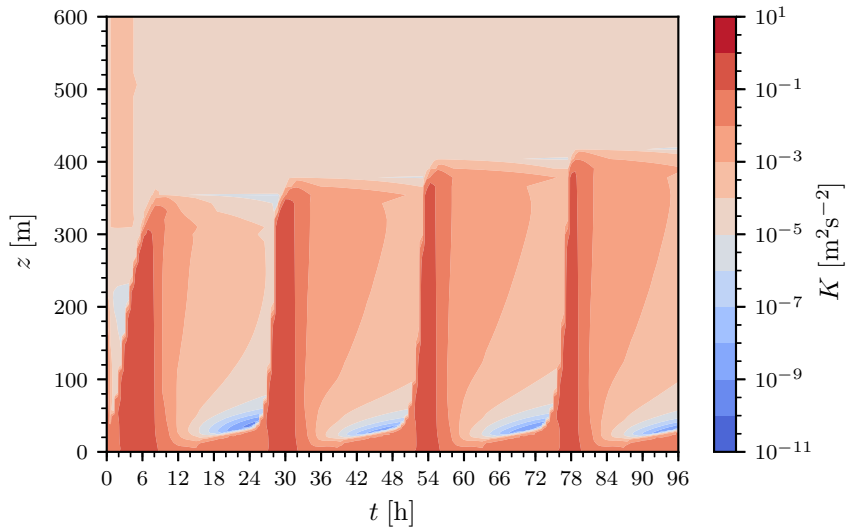}
  \caption{Contour plot showing the results of turbulent kinetic energy in the EARS model as it changes with height and time during the simulation}
  \label{fig:TKE}
\end{figure}

Figure \ref{fig:TKE} shows the evolution of TKE in the domain. As surface temperature rises in the morning, heat is transferred from the ground into the surface layer. Warm air from the surface layer starts rising and enhances turbulent mixing and increases the TKE. This leads to development of the convective ABL. Conversely, when surface temperature decreases the TKE rapidly declines. During the transition from convective to stable ABL turbulence close to the ground dies out faster than turbulence higher up in the mixing layer. This leads to development of a residual layer during nighttime in which the turbulence is slowly decaying, which is natural because of the large length- (and time-) scales there. This is shown in Fig.~\ref{fig:l} where turbulent length scale in the residual layer is two orders of magnitude larger than near the surface. Higher levels of TKE in the residual layer are also observed in Fig.~\ref{fig:TKE}. The maximum height to which turbulence extends in the ABL increases. This is related to an increase of the surface momentum and heat flux maxima. The TKE behaves in a nearly periodic manner after the third day at which time the results are essentially independent of initial conditions. Note that TKE is weak in the region between the residual turbulence and LLJ. The reason for this is related to the negative horizontal wind speed gradient above the core of the LLJ and associated with weak turbulence shear production in that region. Therefore, as the turbulence activity in that region is low the turbulent length scale also becomes small.

\begin{figure}
  \centering
  \includegraphics{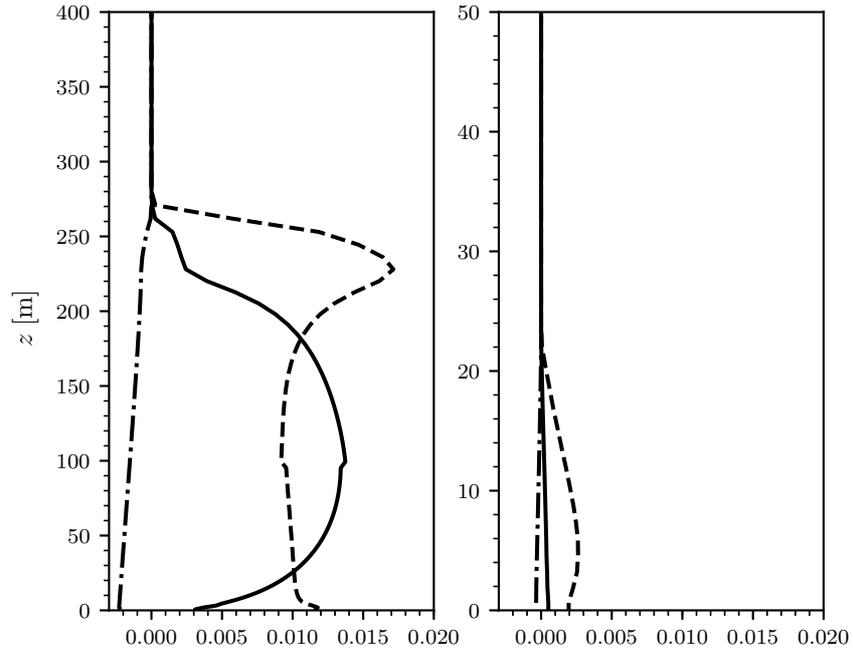}
  \caption{Profiles of scaled turbulent momentum flux at $t = 5 \; \mathrm{h}$ (left) and $t = 17 \; \mathrm{h}$ (right) corresponding to convective and stable periods with sum of horizontal variance $\sqrt{\overline{uu}^2 + \overline{vv}^2}$ (dashed), vertical variance $\overline{ww}$ (full line) and covariance between horizontal and vertical fluctuations $\overline{uw}^T$ (dash-dotted). The turbulent fluxes are scaled with ${\mathbb{V}_g}^2$}
  \label{fig:momentum_fluxes}
\end{figure}

\begin{figure}
  \centering
  \includegraphics{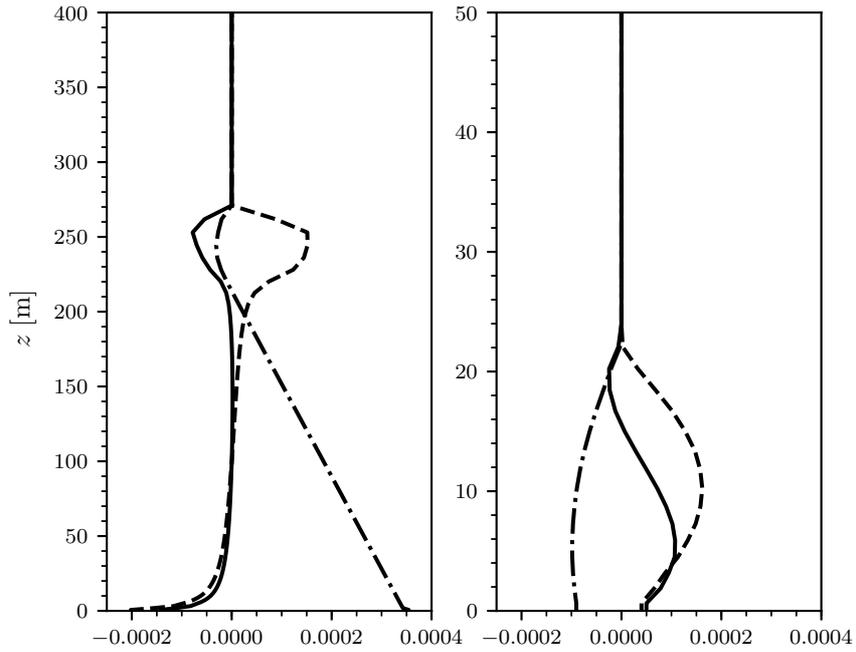}
  \caption{Turbulent heat fluxes along the x- $\overline{u\theta}$ (dashed), y- $\overline{v\theta}$ (full line) and $\overline{w\theta}$ (dash-dotted) directions at $t = 5 \; \mathrm{h}$ (left) and $t = 17 \; \mathrm{h}$ (right) corresponding to convective and stable periods. The turbulent fluxes are scaled with velocity and temperature scales respectively $\mathbb{V}_g$ and $\Delta T = max(\Theta_s) - min(\Theta_s)$}
  \label{fig:heat_fluxes}
\end{figure}

TKE indicates the intensity of turbulence but it does not show the direction in which turbulence is transporting momentum and heat. Turbulent fluxes adjust to thermal stratification of the ABL, which results in TKE being distributed differently under stable and convective conditions. Figures \ref{fig:momentum_fluxes} and \ref{fig:heat_fluxes} show the scaled turbulent momentum and heat flux profiles during the convective and stable periods of the diurnal cycle at $t = 5 \; \mathrm{h}$ and $t = 17 \; \mathrm{h}$, respectively. The scaling parameters for velocity and temperature are $\mathbb{V}_g$ and the difference between maximum and minimum of the surface temperature $\Delta T = max(\Theta_s) - min(\Theta_s)$. The scaling allows comparison between turbulent fluxes in convective and stable periods. Figures \ref{fig:momentum_fluxes} and \ref{fig:heat_fluxes} show that turbulent momentum and heat fluxes are larger in the convective ABL. During this period the vertical mixing is strong, in this well mixed convective region turbulence is not far from isotropy, but near the inversion layer one should expect the horizontal fluctuations to dominate. This is captured by the EARS model as seen in Figs.~\ref{fig:momentum_fluxes} and \ref{fig:heat_fluxes}. In contrast, the EARS model shows that the vertical mixing is strongly damped in the stable phase, during the night, with a very low level of vertical variance $\overline{ww}$. The TKE is then dominated by the horizontal components making the horizontal fluxes dominant. Figure \ref{fig:heat_fluxes} shows that the heat fluxes behave similarly as the momentum fluxes, i.e., vertical turbulent heat fluxes are dominant in the convective ABL and horizontal turbulent heat fluxes are dominant in the stable ABL. The EARS model accounts for turbulence anisotropy unlike the simpler eddy-viscosity turbulence models. Such models also require special modelling of the horizontal turbulent heat fluxes.

\section{Conclusions} \label{sec:conclusion}

We have extended the work of \cite{vzeli2020modelling} and \cite{vzeli2020explicit}, where stably stratified and convective ABLs are simulated with the EARS model, by studying a diurnal cycle of the so-called GABLS4 experiment. Compared to the previous EARS model studies, where the model was validated for quasi-steady ABL, our focus here is to model the transitioning ABL using the same model formulation and coefficients as in the previous studies. The GABLS4 simulation is extended to four consecutive days with periodic temperature forcing at the surface. Compared to the original GABLS4 experiment, the extended diurnal cycle allows us to study transitions from stable to convective as well as from convective to stable ABL. It also reveals that in the initial twentyfour hours the ABL results are still substantially influenced by the specifics of the initial conditions.

The results of the EARS model are compared to several LES reported by \cite{couvreux2020intercomparison}. The results show that the EARS model is able to predict first-order statistics with good accuracy in both convective and stable periods of the day. During the convective part of the day the horizontal wind speed and potential temperature profiles are fairly constant due to strong turbulence mixing. The potential temperature profile in the convective ABL is composed of a lower part with negative gradient and a higher part where potential temperature increases with height. This agrees with observations reported by \cite{vzeli2020explicit} and is the result of the non-gradient contribution in the solution for the turbulent heat flux. During the period of surface cooling the ABL transitions from convective to stable. The transition is followed by a decrease of the ABL height and formation of a LLJ. The profile of horizontal wind speed and LLJ in LES is well predicted by the EARS model. However, it is shifted one hour later.

Aside from the LLJ, the EARS model predicts other physical processes that are related to turbulence mixing. The turbulence lengthscale increases above the stable ABL, at nighttime, which is followed by decaying turbulence from the previously convective ABL. The model also simulates increasing surface heat flux resulting from ground heating and heating by warm air from higher altitudes approaching the surface when turbulence starts to develop. Compared to simpler models, the EARS model is also able to predict the complete Reynolds-stress tensor and heat flux vector, i.e., including the non-zero horizontal turbulent fluxes, which are comparable in magnitude to the vertical fluxes. The present study shows that EARS model has a generalized solution that renders it applicable across varying stability ranges and thermal stratification, and captures the essentials of the transitions between the two types of states.

\section*{Acknowledgment}

The financial support from the Bolin Centre for Climate Research is gratefully acknowledged. Furthermore, Geert Brethouwer is supported by the Swedish Research Council through Grand 621-2016-03533.

\bibliography{ref.bib}
\bibliographystyle{abbrvnat}

\end{document}